\documentclass[a4paper]{jpconf}
\usepackage{graphicx}
\usepackage{mythesis}
\usepackage{hyperref}
\pdfoutput=1
\begin{document}
\title{Beam monitoring and Near detector requirements for a Neutrino factory or long baseline beams }

\author{Rosen Matev, Roumen Tsenov (presenter)}

\address{Department of Atomic Physics, St. Kliment Ohridski University of Sofia, 5 James Bourchier Boulevard, Sofia 1164, Bulgaria}

\ead{tsenov@phys.uni-sofia.bg}

\begin{abstract}
\NuFact is a facility for future precision studies of neutrino oscillations.
A so called near detector is essential for reaching the aimed precision of neutrino oscillation 
analysis.
Main task of the near detector is to measure the flux of the neutrino beam. Such brilliant neutrino 
source like \NuFact provides also opportunity for precision studies of various neutrino interaction processes in the
near detector.
We discuss design concepts of such a detector. Results of simulations of a high resolution scintillating fiber tracker show that it is capable to measure the neutrino flux through pure leptonic interactions with an uncertainty of  the order of 1\percent.
A full set-up of the near detector consisting of high granularity vertex detector, high resolution tracker and muon catcher is also presented.
\end{abstract}

{\it Invited paper to
{\bf NUFACT11}, the XIII-th International Workshop on Neutrino Factories, Super beams and Beta beams, 1-6 August 2011, CERN and University of Geneva. (Submitted to IOP conference series.)}

\section{Neutrino Factory Near Detector(s)  baseline}

A future neutrino facility  \cite{Choubey:2011zz} will need near detectors in order to perform oscillation measurements 
with required sensitivity. It is necessary to have one
near detector for each of the straight sections of the storage ring at each of the two polarities, so four near detectors designed to carry out measurements essential to the oscillation-physics programme are
required.
The near detector tasks  include
 measurement of neutrino flux through the measurement of neutrino-electron scattering;
measurement of neutrino beam properties needed for the flux to be extrapolated to  the far 
  detector;
 measurement of charm production cross sections (charm production in far detectors is one
  of the principal backgrounds to the oscillation signal). In addition, the brilliant Neutrino factory beam allows for  unique neutrino physics non-oscillation studies, such as
measurement of cross sections, structure functions, nuclear effects, $\sin^2 \theta_W$ 
{\it etc.} at neutrino 
energies in the \unit{0-25}{\GeV} range. The near detector must also be capable of searching for new physics, for
example by detecting \tau-leptons which are particularly sensitive probes of non-standard interactions
at source and at detection. $\nu_{\tau}$ detection is also important in a search for sterile neutrinos.

Design requirements for the near detector(s) can be formulated as follows:
 low Z high resolution tracker for flux and cross-section measurement ($\nu_\mu$ and $\nu_e$); 
 magnetic field for better than in MIND
~\cite{Laing}
 muon momentum measurement;
 muon catcher and capability for  \APelectron/\Pelectron identification;
 vertex detector for charmed hadrons and \tau-leptons detection (for non-standard interactions and sterile neutrinos searches);
 good resolution on neutrino energy (much better than in the Far Detector) for flux extrapolation.

\begin{figure}[h]
  \includegraphics[width=25pc]{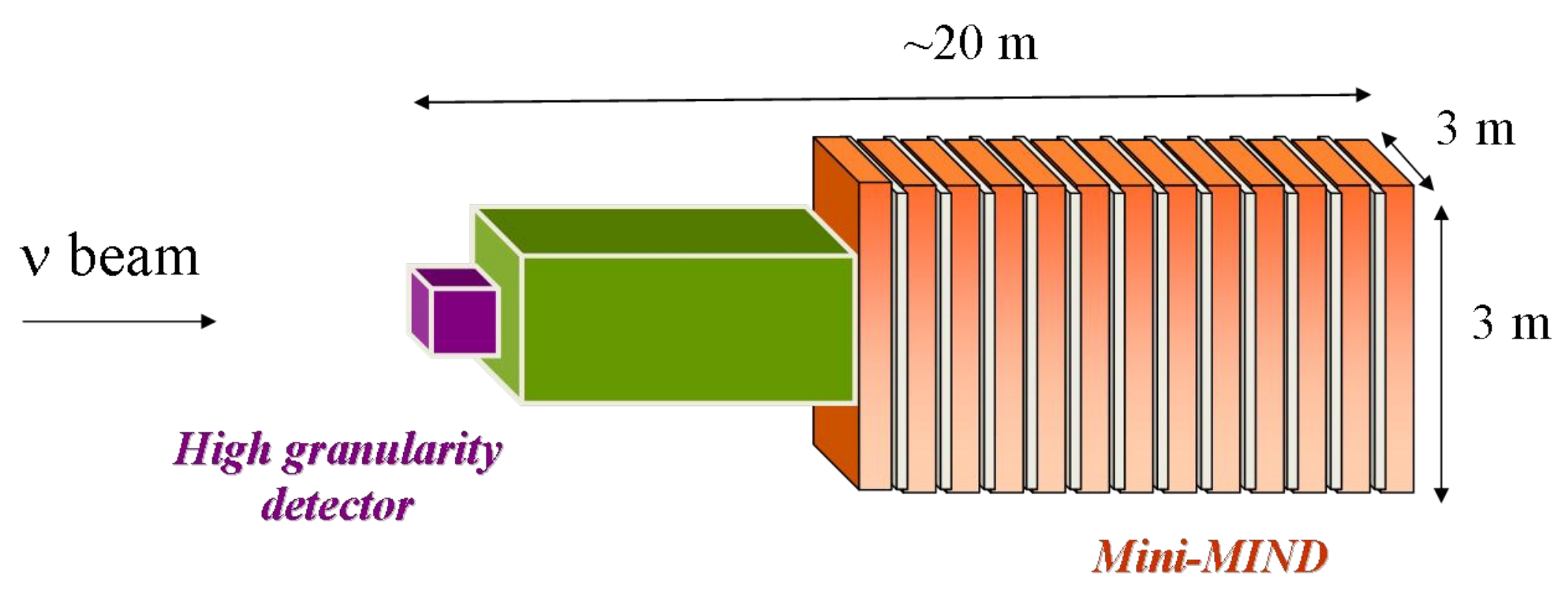}\hspace{1pc}%
\begin{minipage}[b]{10pc}
    \caption{Block diagram design of the Near Detector}
   \label{fig:NDdesign}
\end{minipage}
\end{figure}

Current near detector  design anticipates  three subdetectors (Fig.~\ref{fig:NDdesign}):
high granularity detector for charm/\tau measurement;
 high resolution tracker
  for precise measurement of the event close to the vertex and Mini-MIND detector for muon measurement.

\section{Measurement of the neutrino flux by neutrino-electron scattering}

Neutrino-electron interaction cross sections are straightforwardly calculated in the 
Standard Model framework. Any small uncertainties could come only from (well measured) 
Standard Model parameters. Therefore, such processes are suitable for measurement of neutrino 
beam fluxes, provided that beams are intense enough.

There are two pure leptonic interactions of  the \NuFact beam producing a muon:
\begin{equation}
  \label{eq:imd}
  \HepProcess{\Pnum + \Pelectron \to \Pmuon + \Pnue} \\
  \text{ and} \quad
   \HepProcess{\APnue + \Pelectron \to \Pmuon + \APnum}  \quad \text{(\emph {IMD})} .
\end{equation}
The first one is known as inverse muon decay, while the second one produces muon in the final state through annihilation.
The neutrino energy threshold (for electrons at rest) for both processes is
 $\unit{10.9}{\GeV}$ .
 
\begin{figure}[ht]
\begin{minipage}{17pc}
  \includegraphics[width=\textwidth]{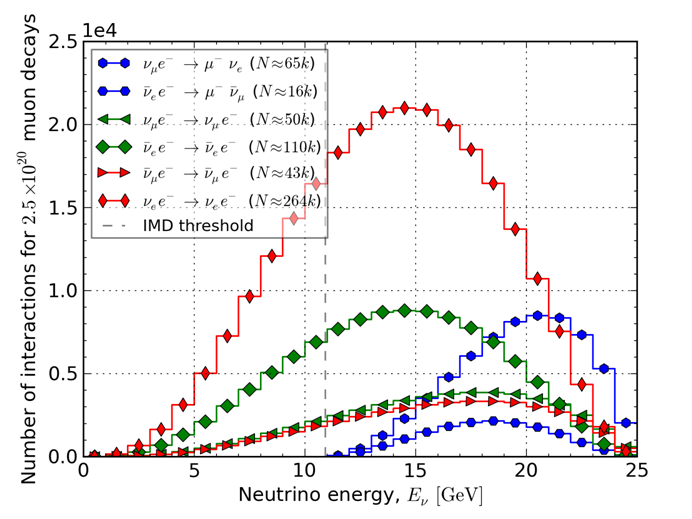}
  \caption{Number of neutrino-electron 
    interactions for
    a nominal year of
     \NuFact operation. Rates are calculated for 
    \unit{2.7}{\ton} 
    detector with \unit{1.5 \times 1.5}{\meter^2} frontal cross section and average $Z/A \approx 
    0.54$. Detector is placed \unit{100}{\meter} after the straight section end. Dashed vertical 
    line indicates threshold for quasi-elastic scattering.}
    \label{fig:eventRate}
    \end{minipage}
      \hspace{2pc}%
    \begin{minipage}{17pc}
  \includegraphics[width=\textwidth]{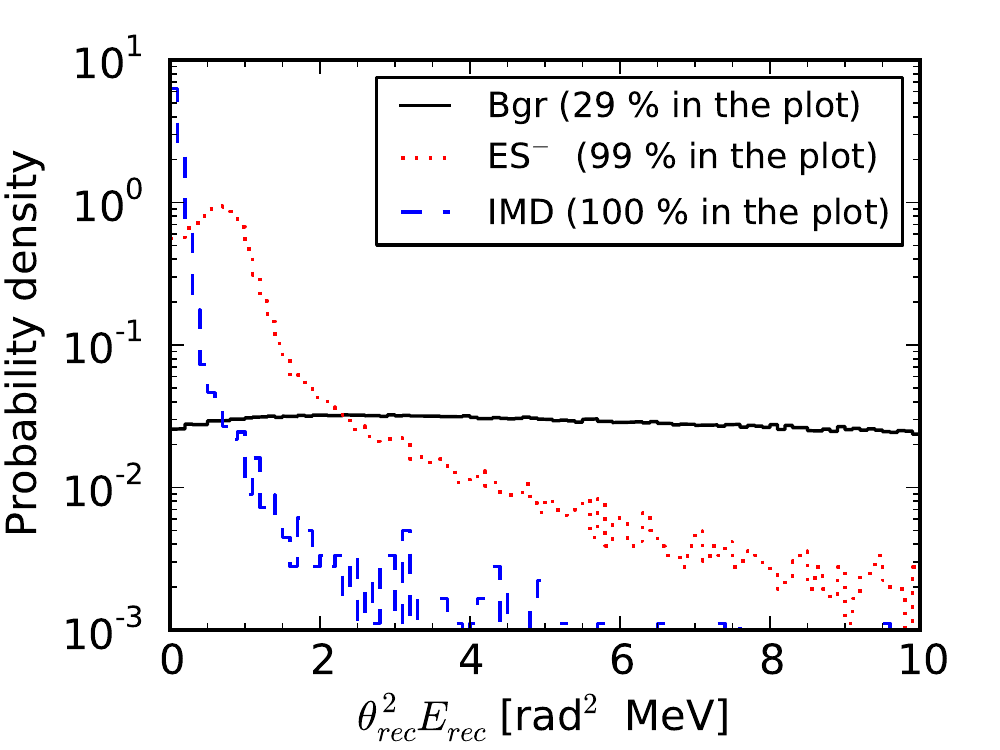}
  \caption{Distributions of reconstructed 
\inelast variable for 
  \IMD (blue), \ESm (red) and background (black) events in \Pmuon--decay mode. The fraction of events 
  contained in the plot is indicated in the legend. All distributions are normalized to a unit 
  area.}
    \label{fig:bgrKin}
    \end{minipage}
  \end{figure}

There are four pure leptonic  reactions of interest producing an energetic electron:
\begin{equation}
  \label{eq:esnum}
  \Pnum + \Pelectron \to \Pnum + \Pelectron \\
  \text{and} \quad 
    \APnue + \Pelectron \to \APnue + \Pelectron  \quad \text{(\emph{\ESm})}
\end{equation}
\begin{equation}
  \label{eq:esnue}
  \Pnue + \Pelectron \to \Pnue + \Pelectron \\
  \text{and} \quad
 \APnum + \Pelectron \to \APnum + \Pelectron \quad \text{(\emph{\ESp})}.
\end{equation}
Despite smallness of the total cross sections for the above processes, massive detector placed close to
the straight section end can provide sufficient interaction rate - \FigureRef{fig:eventRate}. 
However, inclusive CC and NC neutrino interactions with nuclei
\begin{equation}
  \label{eq:ccbgr}
  \Pnulepton + N \to \Plepton + X \\
  \text{ and} \quad
  \Pnulepton + N \to \Pnulepton + X
\end{equation}
are a few orders of magnitude more frequent. For example, 
at \unit{15}{\GeV} the muon neutrino CC total cross section is
$\sim 1 \times 10^{-37} \unit{}{\cm^2}$,
compared to
$\sim 2 \times 10^{-41} \unit{}{\cm^2}$
for inverse muon decay $\Pnum \Pelectron \to \Pmuon \Pnue$.
An obvious distinction between purely leptonic processes and processes 
(\ref{eq:ccbgr}) is the lack of hadronic system X in the former. Thus, measured 
 recoil energy of the hadronic system can be used as a good criterion for background 
suppression.
Muons from quasi-elastic neutrino-electron scattering  (\ref{eq:imd}) have angular distribution peaked at very forward 
direction. At the \NuFact, the polar angle of these muons does not exceed \unit{5}{\mrad}. The 
angular spread comes mainly from the intrinsic 
scattering angle $\sim \unit{4}{\mrad}$ in these processes, while 
neutrino beam divergence (and solid angle covered by detector) makes little contribution. This 
kinematic property can be used as another event selection criterion. Polar angle 
distribution of electrons from neutrino-electron elastic scattering (\ref{eq:esnum}, \ref{eq:esnue}) is ten times wider and is not 
suitable for event selection. On the other hand, the composite variable \inelast ,  proportional to Bjorken's $y=1-E_l/E_{\Pnu}$ 
 in elastic scattering,
provides good separation between signal and background for all neutrino-electron scattering 
processes provided lepton angle and energy are measured with sufficient precision. Two options for the high-resolution tracker are being considered: {\em scintillating fiber tracker} and  {\em straw tube tracker}.

\section{Scintillating fiber tracker}
\begin{figure}[h]
  \includegraphics[width=26pc]{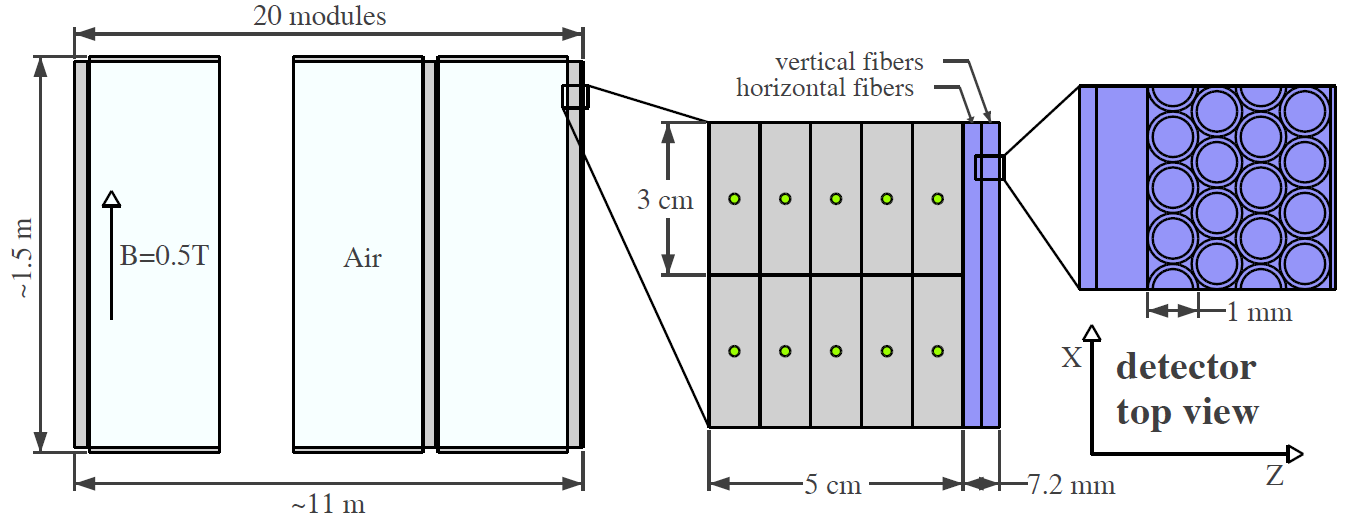}\hspace{1pc}%
\begin{minipage}[b]{10pc}
    \caption{\label{fig:detector}Schematic drawing of the detector.}
\end{minipage}
\end{figure}

A schematic drawing of a scintillating fiber tracker with an incorporated 
calorimeter  is shown on Fig. \ref{fig:detector}. The detector consists of 
20 square shaped modules placed perpendicular to the beam axis. Each module has
a calorimeter section and a tracker section (also called tracker station). Modules are positioned 
equidistantly forming gaps filled with
air. With larger distance between tracker stations, X and Y displacement of hits is increased and 
thus angular resolution improved. The sides of the air gaps are covered with layers of plastic 
scintillating bars. 
The detector is placed in \unit{0.5}{\tesla} dipole magnetic field.
Each station consists of one layer of fibers with horizontal 
orientation and another one with vertical orientation. Each layer has four planes made of 
\unit{1}{\mm} cylindrical  fibers. They form a hexagonal pattern in the layer, thus minimizing dead 
volume. There are 12 000  fibers per station, thus 240 000 fibers in total. Calorimeter sections consist of plastic scintillating bars perpendicular to the magnetic field and arranged in 5  planes in each section. 
Bars are
co-extruded with a wavelength shifting (WLS) fibers inside and have \unit{10}{\mm} by 
\unit{30}{\mm} cross-section. Both tracker fibers and WLS fibers in bars are read from both ends 
by silicon photomultipliers (SiPMs).  
Overall dimensions of the detector are $\sim \unit{1.5}{\meter} \times \unit{1.5}{\meter} \times 
\unit{11}{\meter}$ and the detector mass is $\sim \unit{2.7}{\ton}$.
%

\subsection{Simulation of the detector response and signal extraction}

Neutrino flux at the near detector has been generated by a Monte Carlo simulation of
muon decays along the straight section of the \NuFact decay ring
 \cite{Karadzhov:2009zz,Karadzhov:2010su}. 
Neutrino interactions in the detector have been simulated by the GENIE package
~\cite {Andreopoulos:2009rq}.
For the simulation of the detector response to them, the Geant4 software platform
~\cite{Agostinelli:2002hh}  was used. Simple algorithms have been developed for vertex and scattered lepton track reconstruction.

\begin{figure}\begin{center}
  \includegraphics[width=0.4\textwidth]{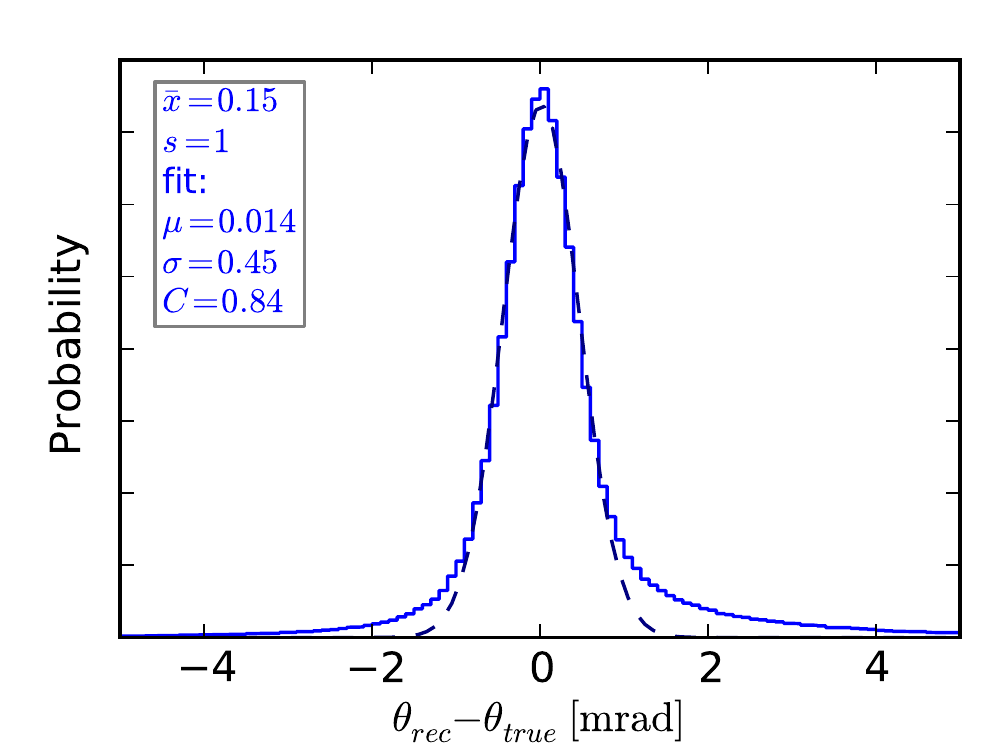}
  \includegraphics[width=0.4\textwidth]{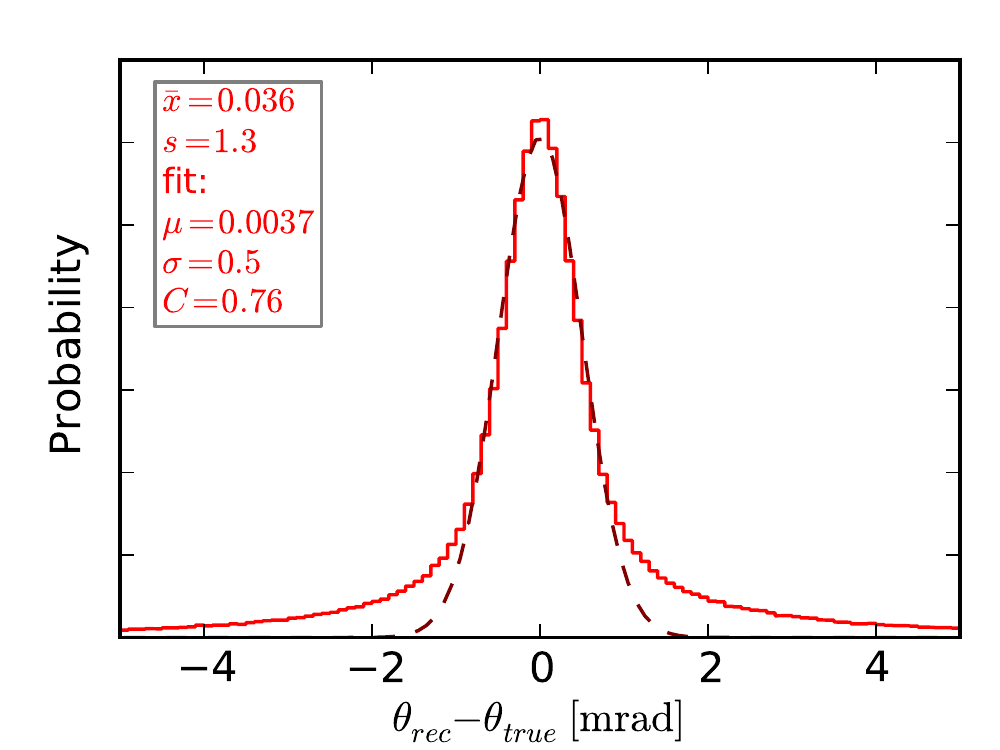}
  \caption[Angular resolution of primary track.]{Obtained angular resolution for muons ({\bf left}) 
and electrons ({\bf right}). Gaussian fits are shown with dashed lines. }
  \label{fig:angularResolution}\end{center}
\end{figure}

\begin{figure}\begin{center}
  \includegraphics[width=0.4\textwidth]{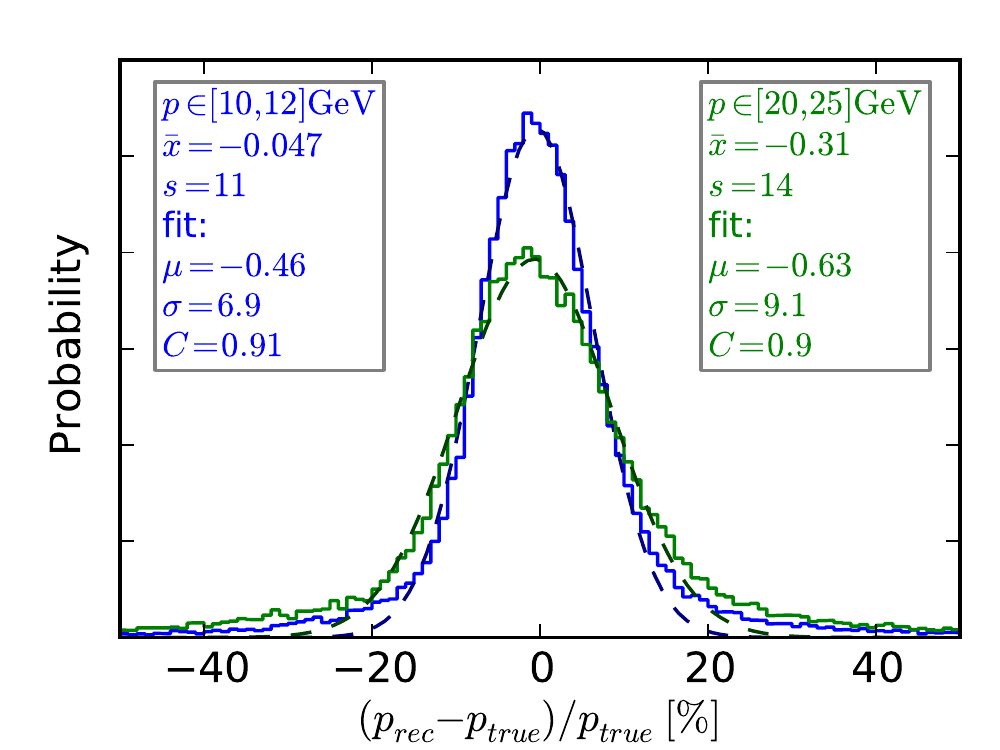}
  \includegraphics[width=0.4\textwidth]{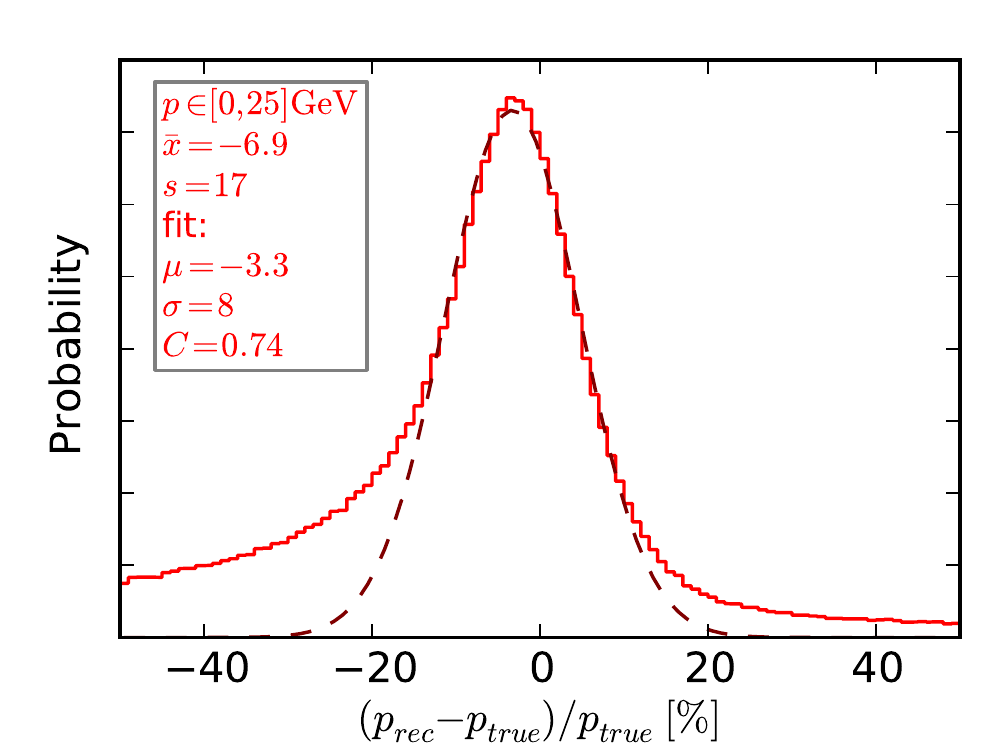}\end{center}
  \caption[Momentum resolution of primary track.]{Obtained momentum resolution for muons ({\bf left}) 
and electrons ({\bf right}). Gaussian fits are shown with dashed lines. For muons, distribution is shown for 
two samples of events: one with true muon momentum in the \unit{[10-12]}{\GeV} range (blue) and 
one in the \unit{[20-25]}{\GeV} range (green).}
  \label{fig:momentumResolution}
\end{figure}

 Distribution of the difference between 
reconstructed and true value of the scattering angle   is shown in 
\FigureRef{fig:angularResolution}.  The 
resolution ($\sigma$ parameter of the fit) is $\sim \unit{0.5}{\mrad}$ for both muons and 
electrons. Reconstructed momentum resolution
is shown on \FigureRef{fig:momentumResolution}. For muons it goes up to 
$\sim \unit{9}{\percent}$ for the highest energy muons. For electrons, the distribution is biased 
towards the negative values with a heavy negative tail. The reason for this  is that they loose momentum due to bremsstrahlung and ionization. If 
energy loss is taken into account, for instance with Kalman filter fitting
~\cite{Kalman:1960}, bias can be reduced. 


Both \IMD and \ES events have a property of low (consistent with single particle) energy 
deposition near the vertex. To exploit that, a cut on energy deposit around the vertex is imposed. 
Some other kinematic and calorimetric cuts have also been applied in order to get sample enriched with signal events.
As a result, signal-to-background ratio has been increased from  $\sim 10^{-4}$  to 
$\sim \unit{30-50}{\percent}$.
Further on, extrapolations of certain background distributions should be made in order to subtract 
background from event samples. 
We have chosen to do background subtraction in terms of primary lepton (\Pmu or \Pe) 
kinematic variables -- lepton's scattering angle \scangle and initial momentum $p$  measured in the 
detector. 
In the case of \IMD signal extraction, scattering angle \scangle and \inelast variable 
can be used to discriminate signal from background. In the case of \ES signal, background is well 
separated only when one exploits \inelast variable. The distributions of  \inelast 
for \IMD signal, \ESm signal and background are shown in 
\FigureRef{fig:bgrKin}. Background distribution  is nearly flat. This fact 
allows for its simple parameterization. 

Two methods of obtaining the number of signal events are discussed below {\emph linear fit method} and  $\mu^+${\emph-method}.

\begin{figure}\begin{center}
  \includegraphics[width=0.4\textwidth]{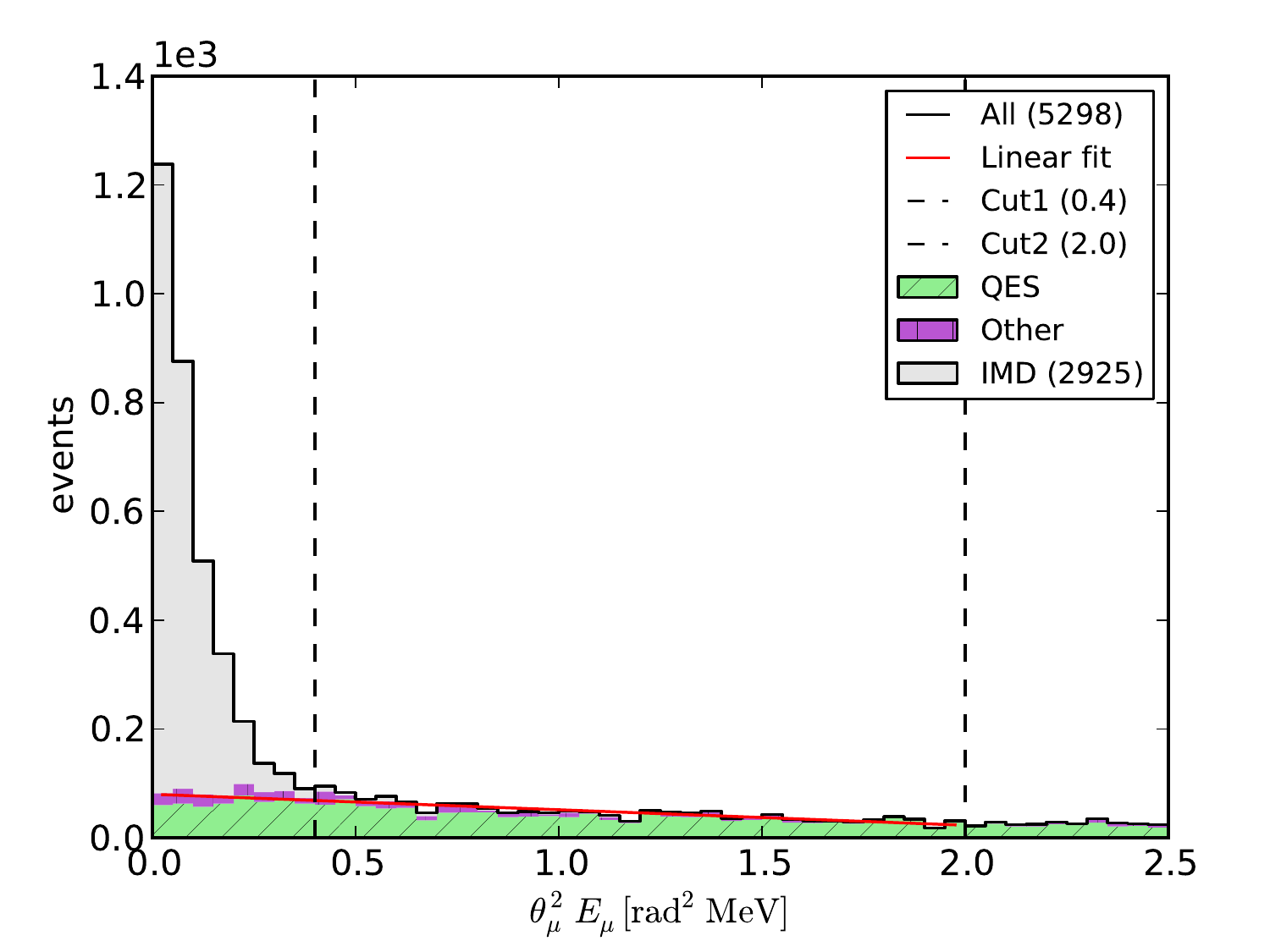}
    \includegraphics[width=0.4\textwidth]{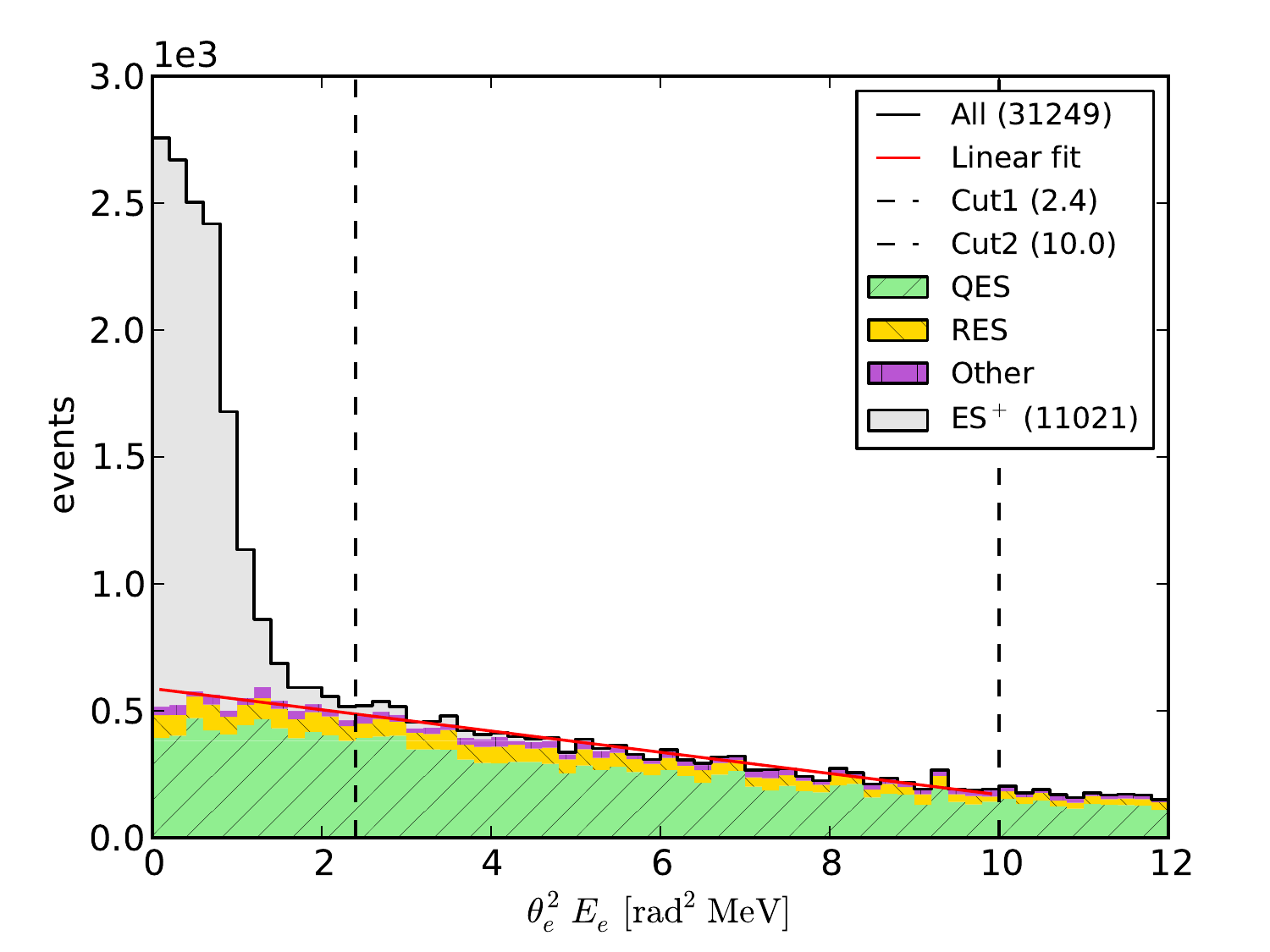}\end{center}
  \caption {Distributions over \inelast for the \IMD sample ({\bf left}) and  for the \ESp sample ({\bf right}).
  The leptonic events histogram is
  filled with solid gray, the hadronic events histogram is hatched and the total spectrum is in 
  black. The two cuts bounding the fit interval are drawn with dashed line. The red line
  indicates the background extrapolation.}
  \label{fig:imdLinear}
\end{figure}

Linear fit method relies on the nearly flat shape of the respective background distribution. The 
idea is to estimate the background under the signal peak by linear extrapolation from 
signal-free region towards the signal one.
Examples
are shown in Fig. \ref{fig:imdLinear}. Comparison 
between the estimated and the true number of signal events is made in \TableRef{tab:fitResults}. 
It is seen that the true values lie within the \unit{95}{\percent} confidence intervals of the 
predictions. 
The systematic uncertainty, estimated as the difference between the fit result and the true number of signal events is of the order of \unit{1}{\percent}. However, to give conclusive estimation of the systematic error, 
one should investigate if and how various parameters of simulation and selections influence the 
background shape.

\begin{figure}\begin{center}
  \includegraphics[width=0.4\linewidth]{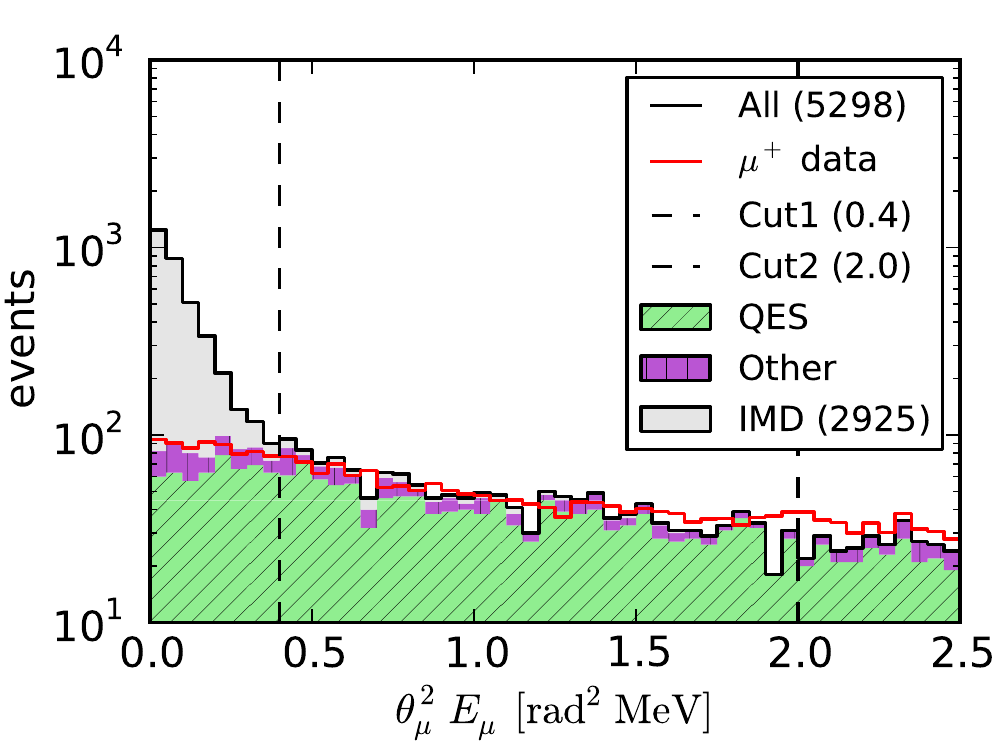}
  \includegraphics[width=0.4\linewidth]{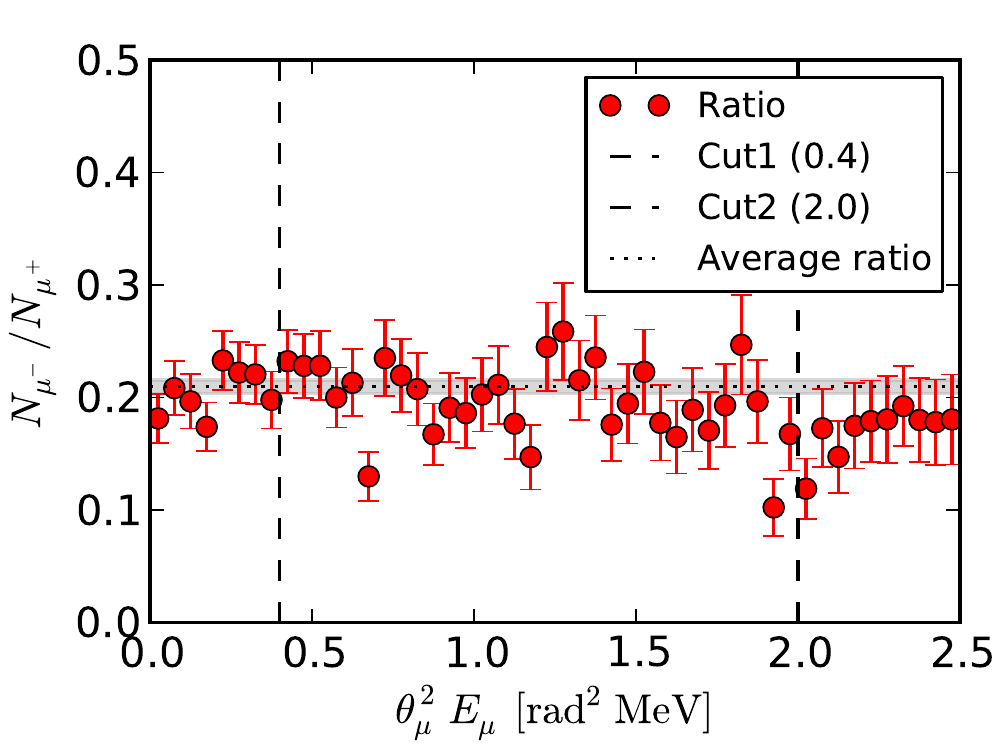}\end{center}
  \caption[Distributions over \inelastMu for the \IMD sample. \APmuon-method for background 
  subtraction.]%
  {{\bf Left:} distributions over \inelastMu for the \IMD sample. The leptonic events 
  histogram is filled with solid gray, the hadronic events histogram is hatched and the total 
  spectrum is in black. The two cuts bounding the normalization interval  are drawn with 
  dashed line. The red line indicates the normalized \APmuon histogram. {\bf Right:} ratio of 
  the \Pmuon histogram and the \APmuon histogram over \inelastMu. Horizontal dotted line indicates the (constant) normalization factor.
  }
  \label{fig:imdMuplus}
\end{figure}

\IMD interactions are present only in the \Pmuon decay mode. The idea of the \APmuon-method is to 
estimate the background under the \IMD signal peak exploiting the distribution of positive muons 
detected in (\APnum, \Pnue)-beam  \cite{Geiregat:1990rb}. In the near detector, an event sample 
from the (\APnum, \Pnue)-beam ({\em i.e. beam with reversed muon polarity}) events is selected with the same selection cuts as the \IMD sample. 
The \inelastMu histogram for \APmuon is normalized to the \inelastMu 
histogram for \Pmuon.  An interval outside the \IMD signal peak and with approximately 
constant ratio of \Pmuon- and \APmuon-events is defined and normalization factor is calculated within this interval.
The \Pmuon histogram and the normalized \APmuon histogram are 
shown in \FigureRef{fig:imdMuplus} (left).  Subtraction is made using normalised \APmuon histogram.

\begin{table}
  \caption[Estimated number of signal events.]%
    {Estimated number of signal events for the three event samples. The result in the last row was 
      obtained using the \APmuon background subtraction method, while the other three results were 
      obtained using linear fit background subtraction method.
      Statistics correspond to $2.3\times 10^{19}$ \Pmuon decays and $2.3\times 10^{19}$ \APmuon 
      decays, which is approximately a tenth of the nominal year.}
  \vspace{2ex}
  \begin{tabular}{lcccccc}
  \br
    Event  & Selection & Overall & Purity & All events & Signal events & Signal events \\
    sample & eff.      & eff.    &        &            &               & from fit      \\
    \mr
    \IMD         & \unit{86}{\percent} & \unit{46}{\percent} & \unit{81}{\percent} &
      3520 & 2850 & 2926 $\pm$ 59 \\
    \ESm         & \unit{70}{\percent} & \unit{32}{\percent} & \unit{61}{\percent} &
      7355 & 4491 & 4479 $\pm$ 86 \\
    \ESp         & \unit{83}{\percent} & \unit{37}{\percent} & \unit{63}{\percent} &
      16964 & 10607 & 10512 $\pm$ 131 \\
    \mr
    \IMD         & \unit{86}{\percent} & \unit{46}{\percent} & \unit{81}{\percent} &
     3520 & 2850 & 2831 $\pm$ 61 \\
      \br
      \end{tabular}

  \label{tab:fitResults}
\end{table}

 \TableRef{tab:fitResults} demonstrates that the number of 
neutrino-electron scattering events can be measured exploiting the \inelast distribution with a good precision. A direct 
comparison between measured and true number of signal events shows a deviation of the order of 
\unit{1}{\percent}. It is worth noting that MC truth was not used in reconstruction and signal 
extraction.  Thus, with the presented design of the tracker we can achieve ~1\percent uncertainty on the flux normalisation by exploring IMD and/or ES scattering.

\section {High resolution straw tube tracker}
Another option for the near detector is a high resolution straw tube tracker inspired on the HiResM$\nu$  detector  \cite{Mishra:2008nx} being considered for the LBNE project at Fermilab 
 as a near detector \cite{LBNE}.

Building upon the NOMAD-experience  \cite{Altegoer:1997gv}, this low-density tracking detector will have a fiducial mass of 7.4 tons as an active neutrino target, similar to the ATLAS Transition Radiation Tracker  \cite{Akesson:2004nj} and the COMPASS detector  \cite{Bychkov:2002xw}. The tracker will be composed of straw tubes with 1 cm diameter, in the vertical ($y$) and horizontal ($x$) direction. In
front of each module a plastic radiator made of many thin foils allows the identification of electrons
through their transition radiation. The nominal fiducial volume is:
$350\times 350\times 600$~cm$^3$, corresponding to 7.4 tons of mass with an overall density $\rho < 0.1$~g/cm$^3$.  
The straw-tube tracker will be surrounded by an electromagnetic calorimeter (sampling Pb/scintillator) covering
the forward and side regions. Both sub-detectors will be installed inside a dipole magnet providing
a magnetic field of $\sim 0.4$~T. 

\begin{figure}[h]
 \includegraphics[width=22pc]{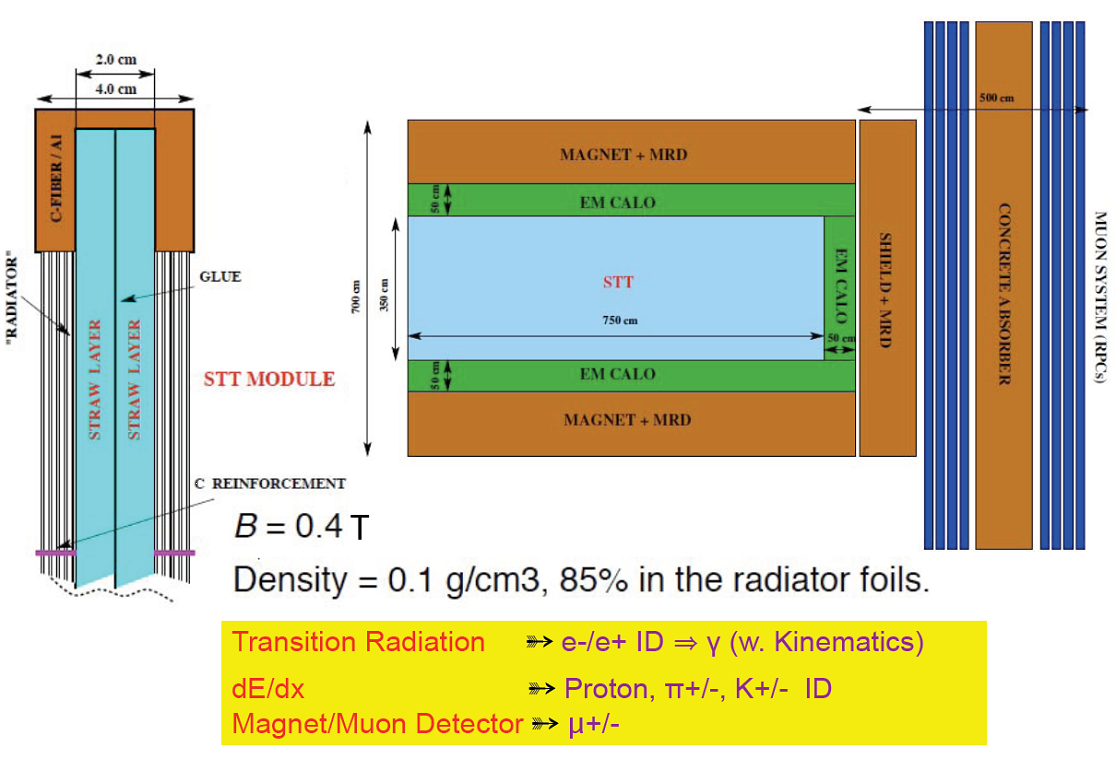}\hspace{1pc}%
\begin{minipage}[b]{15pc}
\caption{Sketch of the proposed HiRes detector showing the inner straw tube tracker (STT), the electromagnetic calorimeter (EM CALO) and the magnet with the muon range detector (MRD). Also shown is one module of the proposed straw tube tracker (STT). Two planes of straw tubes are glued together and held by an aluminium frame.}
\label{fig:HiRes}
\end{minipage}
\end{figure}

The detector will provide
full reconstruction of charged particles and $\gamma$s;
identification of electrons, pions, kaons, and protons from $dE/dx$;
electron (positron) identification from transition radiation ($\gamma > 1000$);
full reconstruction and identification of recoil protons down to momenta of 250 MeV;
reconstruction of electrons down to momenta of 80 MeV from curvature in the B-field.

Detailed simulations of this detector have been carried out in the context of the LBNE proposals  \cite{LBNE}. These simulations are to be adapted to the neutrino spectra at a Neutrino Factory to derive the performance parameters of this detector in this context. 

\section{Charm and Tau Detector} 
A near detector at a neutrino factory needs to  measure the charm cross-section to validate the size of the charm background in the far detector, since this is the main background to the wrong-sign muon signature  The charm cross-section and branching fractions are poorly known, especially close to threshold. For this reason, it is paramount to make an independent near detector measurement of the charm cross-section  and make the error in the charm cross-section negligible in the estimation of the neutrino oscillation background.

Since events with \tau-lepton in the final state have a similar signature to charm events, any detector that can measure charm should be able to measure \tau's as well. This is important to explore couplings of Non Standard Interactions (NSI) at source $\epsilon_{\tau\mu}^s$, $\epsilon_{\tau e}^s$ or detection $\epsilon_{\tau\mu}^d$, $\epsilon_{\tau e}^d$.
A semiconductor vertex detector for charm and \tau-lepton detection could potentially be used for this purpose. 
The advantage of this type of detector is that it is able to operate at a high event rate and still have very good spatial resolution. The latter is necessary to distinguish the primary neutrino interaction vertex from the secondary vertex due to the short lived charm hadron or the \tau-lepton. 
The vertex detector could be similar to the NOMAD--STAR detector that was installed for some time upstream of the first drift 
chamber of the NOMAD neutrino oscillation experiment  \cite{Altegoer:1997gv} and was used to measure the impact parameter 
and double vertex resolution to determine the charm detection efficiency. 
The reconstruction of \tau-leptons from an impact parameter signature with a dedicated silicon vertex detector was studied in the NAUSICAA proposal  \cite{GomezCadenas:1995ij}. A  silicon vertex detector with a $B_4 C$ target was proposed as an ideal medium to identify \tau-leptons. Standard $\nu_\mu$ CC interactions have an impact parameter  {\em r.m.s} of 28~$\mu$m, while tau decays have an impact parameter  {\em r.m.s} of 62~$\mu$m. By performing a cut on the impact parameter significance ($\sigma_{IP}/IP$) one can separate  one prong decays of the tau from the background. For three prong decays of the tau, a double vertex signature is used to separate signal from background. The total net efficiency of the tau signal in NAUSICAA was found to be 12\%. 

A silicon strip vertex detector as part of the near detector could have the following dimensions  \cite{Choubey:2011zz}:
18 layers of  $B_{4}C  (2.49 g/cm^3)$, 150x150x2$cm^3$ each;  total mass = 2.02 \ton,
20 layers of silicon strip or pixel detectors, e.g. $45 m^2$ of silicon; 
about 64 000 channels per layer, 1.28 million channels in total. At the \NuFact in such a detector about $3\times 10^7 \nu_{\mu}$ CC interactions per year are expected and $10^6$ charm events among them. With the  efficiency of the tau detection found in NAUSICAA, one could have a sensitivity of  $P_{\mu\tau}< 3\times 10^{-6}$ at 90\% C.L. on the $\nu_{\mu}-\nu_{\tau}$ conversion probability. 

\section{Summary and outlook}

Near detector at the Neutrino factory is a valuable tool for neutrino flux measurement and (non-)standard neutrino interactions study. The envisioned set-up consists of high granularity vertex detector followed by high resolution tracker and muon catcher. Silicon vertex detector+SciFi tracker+mini-MIND set-up is most advanced with respect to simulations with \NuFact beam. They show that the neutrino flux can be measured with ~1\percent uncertainty. 
Second option exists for the tracker -- HiResM\nu. Simulations with \NuFact beam are needed to confirm its ability to select and measure neutrino-electron scattering.

Further tasks  include simulation of the full set-up in order to estimate 
systematic errors coming from near-to-far extrapolation (determination of the so-called {\it migration matrices});
expectations on cross-section measurements and  
other physics studies, 
sensitivity to non-standard interactions (\tau-lepton production). In a bit more distant future serious
R\&D efforts would be needed to validate the technology choices for the vertex detector, high-resolution tracker, {\it etc.}

\end{document}